\documentstyle[preprint,aps,epsfig]{revtex}

\tighten
\begin{document}
\def\eqn#1{Eq.$\,$#1}
\draft
\preprint{}
\title{Analytical results for random walk persistence}

\author{Cl\'ement Sire$^1$, Satya N. Majumdar$^2$ and Andreas
R\"udinger$^{3,1}$}
\address{$^1$ Laboratoire de Physique Quantique (UMR C5626 du CNRS),
Universit\'e Paul Sabatier
31062 Toulouse Cedex, France (clement@irsamc2.ups-tlse.fr)\\
$^2$ Tata Institute of Fundamental Research, Homi Bhabha Road,
Mumbai-400005, India\\
$^3$ Institut f\"ur Theoretische und Angewandte
Physik, Pfaffenwaldring 57, 70550 Stuttgart, Germany}

\maketitle  
\begin{abstract} 
In this paper, we present the detailed calculation of the persistence exponent
$\theta$ for a nearly-Markovian Gaussian process $X(t)$, a problem initially
introduced in [{\sl Phys. Rev. Lett.} {\bf 77}, 1420 (1996)], describing the
probability that the walker never crosses the origin. Resummed
perturbative and non-perturbative expressions for $\theta$ are derived, which
suggest a connection with the result of the alternative independent interval
approximation (IIA). The perturbation theory is extended to the calculation of
$\theta$ for non-Gaussian processes, by making a strong connection between the
problem of persistence and the calculation of the energy eigenfunctions of a
quantum mechanical problem. Finally, we give perturbative and non-perturbative
expressions for the persistence exponent $\theta(X_0)$, describing the
probability that the process remains bigger than $X_0\times\sqrt{\langle
X^2(t)\rangle}$.

\end{abstract}
\vspace{1cm}


\narrowtext

\newpage
\section{Introduction}

A natural quantity that characterizes a given stochastic process $X(t)$ is
its persistence $P(t)$, {\it i.e.}, the probability that this signal has kept
the same sign up to time $t$. For a large class of physical systems (to be
defined more precisely below), persistence decays as a power-law in time,
$P(t)\sim t^{-\theta}$ for large $t$, thus defining
the persistence exponent $\theta$.

This exponent has been studied in experimental systems (breath figures
\cite{breath}, a liquid crystal system mimicking the 2$d$ Ising model
\cite{lq}, soap bubbles \cite{soap}...), and by theoretical means through the
exact solution of models \cite{{AB1},{BD1},{D1},{voter}}, numerical
simulations \cite{{sta},{BD2}}, and general theoretical methods
\cite{{pert},{oerding},{MSBC},{dornic},{newman},{MB}}.

Most theoretical methods restrict themselves to the study of persistence of
stochastic processes that are Gaussian. This is partly because Gaussian
processes are abundant and simpler. Moreover, in many physical situations, the
study of persistence of non-Gaussian signals can be effectively reduced to
that of Gaussian signals\cite{{pert},{MSBC}}. Thus, given a Gaussian process
$X(t)$ of zero mean, the basic question is: what is the probability that it
remains, say, positive up to time $t$? This is a difficult problem that has
been studied by  mathematicians for a long time\cite{BL}. Recently, however, it
has created much interest among physicists.

One of the general methods recently introduced to tackle this difficult
problem, namely the independent interval approximation (IIA) \cite{{MSBC}},
assumes that the interval lengths between zeros of the process $X(t)$ are
statistically independent. This sole assumption permits the closure of a
hierarchy of equations leading to an approximate expression of $\theta$. This
approximation gives very good results  for smooth Gaussian processes ({\it
i.e.} processes with a finite density of zero crossings). Unfortunately, it is
not clear how this assumption can be justified, and whether a Gaussian process
can be said {\it a priori} to be well described by this approximation. 

An approximation for the distribution of the time-integrated ``magnetization''
\cite{dornic,newman}, $M(t)=t^{-1}\int_0^t{\rm sign}(X(u))\,du$, also leads to
good quantitative results for smooth processes, but suffers from the same
conceptual problems as the IIA, and is not even guided by physical intuition.
Still, the study of this quantity has lead to the introduction of a new
quantity, the generalized persistence \cite{dornic}, that is the probability
that $M(t)$ remains above a certain level $M_0$. This quantity which decays
with a persistence exponent depending continuously on $M_0$ has been studied
in the framework of spin systems and for random walkers \cite{dornic}.

Finally, a  systematic $\epsilon$-expansion, which is exact order by order,
has been developed recently for smooth Gaussian processes \cite{MB}. 

However, all these approximate and exact techniques fail for processes that
are singular, that is for which the density of zero crossings is infinite. 
These processes appear in many physical situations such as nearly
Markovian random walkers \cite{pert} or interface growth \cite{krug}.

In this paper, we come back to the first general method proposed, that is
perturbation theory around a Gaussian and Markovian process \cite{pert}. After
introducing the principle of this method (section {\bf II} to {\bf IV}), which
shows a deep connection between the problems of persistence and the energy
spectrum of a quantum mechanical problem, we present a symmetry argument for
the exact form of $\theta$ which leads to more general results for the
persistence exponent (section {\bf V}). These results also reveal a connection
between IIA and perturbative approaches. In section {\bf VI}, we extend the
perturbative approach to the case of non-Gaussian processes further
reinforcing the link with standard quantum mechanics. In section {\bf VII}, we
show that the various approaches introduced can be applied to the computation
of the probability that the signal $X(t)/\sqrt{\langle X^2(t)\rangle}$ remains
higher than any given non-zero-level $X_0$ (generalized persistence). Finally,
in section {\bf VIII}, we illustrate some of the results obtained in the
preceding sections by means of numerical simulations.

\section{Importance of Gaussian stationary processes}

The most popular examples of persistent systems have been taken from the field
of coarsening dynamics \cite{AB}. For instance, let us consider an Ising
spin system after a quench at very low temperature from a high temperature
disordered state. Domains of positive (essentially $+1$) and negative
(essentially $-1$) magnetization grow with a time-dependent typical length
scale $L(t)\sim t^{1/2}$. For this system, the spin persistence, that is, the
probability that a spin has never changed sign, or has never been crossed by
an interface, is known to decay as $t^{-\theta}$, with $\theta=3/8$ in $d=1$
\cite{{AB1}}, and $\theta\approx 0.22$  in $d=2$
\cite{{sta},{BD2},{pert}}.

Due to dynamical scaling the two-time spin correlation function only depends
on the dimensionless ratio of $L$ at both considered times:
\begin{equation}
\langle S(t)S(t') \rangle=f(L(t)/L(t')).
\end{equation} 
This property will be characteristic of a coarsening system and only relies on
the existence of a unique dynamical length scale and the dynamical scaling
hypothesis \cite{AB}. Now, if $L(t)$ behaves as a power-law for large times,
all two-point correlation functions are then functions of $t/t'$. By
considering $\tau=\log(t)$, these correlation functions are then functions of
$\exp|\tau-\tau'|$, or more simply $|\tau-\tau'|$, so that they become
stationary in the fictitious time $\tau$. 

Moreover, in many physical systems \cite{{pert},{MSBC},{crit}}, the question
of computing the persistence for the original dynamical variable ($S(t)$ in
the above example) can be reduced to the study of the persistence of a Gaussian
variable $X(t)$. One possibility is, of course, that the physical variable is
a Gaussian variable itself: this occurs in the study of the persistence of the
diffusion equation \cite{MSBC}, and for the total magnetization persistence of
a spin system quenched at $T<T_c$ \cite{block}, or $T=T_c$ (in the latter
case, the persistence exponent is a new critical exponent \cite{crit}). But in
some other cases, including the Ising and more generally $O(N)$ spin systems,
the original persistence can be shown to be very close to that of a true
Gaussian process $X(t)$. For instance, a local spin in an Ising system behaves
essentially as the sign of such a Gaussian process, $S(t)\approx {\rm
sign}(X(t))$, an important result which was first used within the OJK theory
\cite{{OJK},{AB}}, and later more precisely formalized by Mazenko and
co-workers \cite{{M1},{AB}}. To summarize, we underline the special role
played by Gaussian processes, and will thus restrict our study to this kind of
process.

The next important remark is that if the persistence of the considered
Gaussian process $X(t)$ decays as $t^{-\theta}$, the persistence in terms of
the fictitious time $\tau$ (for which this process is stationary) is expected
to decay exponentially as $\exp(-\theta\tau)$.  Thus, in the following we
restrict ourselves to the study of persistence for a stationary Gaussian process
$X(\tau)$ \cite{{vK},{CH},{Fell},{SC}}. Note that if $L(t)$ does not behave as
a power-law of time, the persistence still decays as a power-law of
$L(t)$ as soon as dynamical scaling is satisfied, and the proper fictitious
time is simply $\tau=\log L(t)$, for which the process $X(\tau)$ is again
stationary.

The most general equation of motion for a stationary Gaussian walker reads,
\begin{equation}
X'(\tau)=-{\lambda} X(\tau) +\int_{-\infty}^\tau J(\tau-\tau')
\eta(\tau')d\tau',
\label{motion}
\end{equation}
where $\eta(\tau)$ is a Gaussian white noise satisfying $\langle\eta(\tau)
\eta(\tau')\rangle=\delta(\tau-\tau')$. Indeed, this equation must be linear
to preserve the Gaussian property, and the coefficient $\lambda$ of $X(\tau)$
must be constant to preserve stationarity. The last term of
\eqn{(\ref{motion})} accounts for memory effects, involving a memory kernel
$J$, and must take the form of a convolution product, again to preserve
stationarity and the Gaussian property (linearity). Note that it is not
necessary to involve higher derivatives of $X$ in this equation of motion, as
they can be accounted for  by a proper choice of the kernel $J$ (see
\eqn{(\ref{cor})} below).

The Markovian case is associated with $J(\tau)=\delta(\tau)$ (no memory
effects), so that the equation of motion becomes,
\begin{equation}
X'(\tau)=-{\lambda} X(\tau) +\eta(\tau).
\end{equation}
The velocity $X'(\tau)$ only involves the noise at the same time $\tau$.
For such a Langevin walker, the two-point correlation function is simply,
\begin{equation}
\langle X(\tau) X(\tau')\rangle=f(\tau-\tau'),\qquad f(\tau)=
\frac{\exp-\lambda|\tau|}{2\lambda}.
\end{equation}
For convenience, the correlator (and the variable $X$) has been normalized
such that $f'(0^\pm)=\mp 1/2$, and, from now on, this will be assumed for all
correlators. This will ensure that,
\begin{equation}
\omega^2\hat f(\omega)\to 1,\quad {\rm when }\quad\omega\to\pm\infty.
\label{deriv}
\end{equation}
Also note that this Markovian correlator $f$ has a cusp at the origin. We
will define a {\it {nearly}} Markovian Gaussian process as one with
a correlator which satisfy the above condition (5). 

In general, the knowledge of the two-point correlation function $f(\tau)$ is
equivalent to that of the equation of motion, as the Fourier transform of $f$
satisfies,
\begin{equation}
\hat f(\omega)=\langle \hat X(\omega) \hat X(-\omega)\rangle=\frac{|
\hat J(\omega)|^2} {\omega^2+\lambda^2}.
\label{cor}
\end{equation}
This actually shows that any correlator $\hat f(\omega)$ can be reproduced by
a proper (not unique) choice of the memory kernel $\hat J$.

In sections {\bf III} and {\bf IV}, we will give a more extensive account of
the perturbative expansion for $\theta$, in the case of a nearly Markovian
Gaussian stationary process, a calculation which was first introduced in
\cite{pert}, and then reproduced in a real time formalism in \cite{oerding}.
This will be followed (section {\bf IV}) by a resummation of this perturbation
theory using a general symmetry argument, and the discovery of an intimate
connection between the IIA and perturbative methods. A new non-perturbative
expression for $\theta$ is also presented, which happens to reproduce
quantitatively most numerical results (section {\bf VIII}).

\section{Persistence: the Markovian case}

Let us now move to the problem of persistence. The probability that a given
walker remains on, say, the positive side of $0$ at all times between 0
and $\beta$ is,
\begin{equation}
P(\beta)=\frac{\int_{X>0}{\cal D}X(\tau)\exp [-{\cal S}] }
{\int {\cal D} X(\tau) \exp [-{\cal S}]}=\frac{Z_1}{Z_0},
\label{Z}
\end{equation}
where,
\begin{equation}
{\cal S}(\beta,\lbrace X(\tau)\rbrace)={{1}\over {2}} \int_0^{\beta}
\int_0^{\beta} X({\tau}_1) g({\tau}_1-{\tau}_2) X({\tau}_2)\,d\tau_1
d\tau_2, 
\label{weight}
\end{equation}
is the Gaussian weight associated with the trajectory $X(\tau)$, and
$g(\tau_1-\tau_2)$ is the inverse of the correlation matrix
$f(\tau_1-\tau_2)$. ${\theta}$ is then calculated from $P(\beta)$ by taking
the limit, 
\begin{equation}
{\theta}=-\lim_{\beta \to +\infty}
{{\beta}^{-1}} \log P(\beta).
\end{equation} 
We can impose periodic boundary conditions for the walker trajectories,
$X(0)=X(\beta)$,  which should not affect the value of $\theta$ in the limit
of large $\beta$. Indeed, in practice, the process will have a finite typical
correlation time, equal to $\lambda^{-1}$ in the example of the Markovian
walker, so that this extra constraint cannot affect the large time persistence
regime.

The path integrals of \eqn{(\ref{Z})} strongly suggest the connection of this
problem to Feynmann integrals in quantum mechanics or statistical field
theory. Let us make this connection more precise. Because of the
periodicity of the trajectories,
the Gaussian weight in \eqn{(\ref{Z})} can
be also written,
\begin{equation}
{\cal S}=\frac{1}{2\beta}\sum_{n=0}^{+\infty} 
\hat g(\omega_n)|\hat X(\omega_n)|^2, 
\end{equation}
where $\hat g(\omega_n)=1/\hat f(\omega_n)$ (the kernel in the expression of
${\cal S}$ is diagonal in Fourier space) and ${\omega}_n={2\pi n/{\beta}}$ are
Matsubara frequencies. First consider a Markovian process for which $\hat
g(\omega)= \omega^2+{\lambda}^2$ (the Fourier transform of
$f(\tau)=\exp(-\lambda|\tau|)/{2\lambda}$ is $[\omega^2+{\lambda}^2]^{-1}$).
${\cal S}$ can be alternatively written as,
\begin{equation}
{\cal S}=\frac{1}{2}\int_0^\beta \left[
\left(\frac{dX}{d\tau}\right)^2+\lambda^2 X^2\right]\,d\tau.
\label{Smarkov}
\end{equation}
We recognize the action in imaginary time ($\beta$ is then an inverse
temperature) of an harmonic oscillator of frequency $\lambda$. The periodicity
of the paths ensures that $Z_0=$Tr$\,[\exp{-\beta H_0}]$ is then the partition
function of an harmonic oscillator, and $Z_1=$Tr$\,[\exp{-\beta H_1}]$, is the
partition function of the same harmonic oscillator, but with an infinite wall
at the origin (as the particle is constrained to remain on the positive axis).
For large time, the persistence behaves as,
\begin{equation}
P(\beta)\sim\exp[-\beta(E_1-E_0)],
\end{equation}
where $E_0$ and $E_1$ are the ground state energies of these quantum systems.
By direct identification, we thus find that, 
\begin{equation}
\theta=E_1-E_0. 
\end{equation} 
Moreover, $E_0=\lambda/2$, and it is easy to convince oneself that the ground
state wavefunction of $H_1$, is the first excited state of $H_0$
restricted to
the positive axis, so that $E_1=3\lambda/2$ (this argument is very general,
and only relies on the $x\to -x$ symmetry of the potential). We finally find
that $\theta=\lambda$ for a Markovian process. This is a well-known fact
\cite{{vK},{CH},{Fell},{SC}}, that can be simply illustrated for the usual
Langevin Markovian walker, for which the equation of motion reads (in actual
time $t$), $\frac{dx}{dt}=\eta(t)$. For such a random walk, the persistence
exponent is known to be $1/2$ \cite{{vK},{CH},{Fell},{SC}}. Let us reproduce
this result within our approach. The two-point correlation function is easily
computed: $\langle x(t)x(t')\rangle= \min(t,t')$, and the normalized variable
$X(t)=x(t)/\sqrt{\langle x(t)^2\rangle}$, has a correlator, $\langle
X(t)X(t')\rangle= (t'/t)^{1/2}$, for $t\geq t'$. This correlator is a function
of the ratio of the two times, so that it is stationary after the change of
variable $\tau=\log(t)$, becoming $\langle X(\tau)X(\tau')\rangle=
\exp\left[-\frac{1}{2}|\tau-\tau'|\right]$. Applying the above calculation, we
recover the result, $\theta=\lambda=1/2$.

\section{Perturbation around a Gaussian Markovian process}

Of course, this heavy machinery is not introduced to deal with the well
understood Markovian case, but rather to be applied  to the case of a nearly
Markovian walker, for which no result exists. Thus, let us consider such a
walker for which,
\begin{equation}
f(\tau)=\frac{1}{2\lambda}\left[\exp(-\lambda|\tau|)+\phi(\tau)\right],
\end{equation}
where $\phi(\tau)$ is assumed to be a ``small perturbation'' to the Markovian
correlator. In Fourier space this can be written, to first order in
$\phi$,
\begin{equation}
\hat g(\omega)=\hat f(\omega)^{-1}=\omega^2+\lambda^2-\hat h(\omega),\qquad
\hat h(\omega)=\frac{(\omega^2+\lambda^2)^2}{2\lambda}\hat\phi(\omega).
\label{h}
\end{equation}
In the general case, the denominator $Z_0$ of \eqn{(\ref{Z})} can be exactly
computed, as any unconstrained Gaussian integral, and is proportional to ${\rm
det}^{1/2}[f(\tau_i-\tau_j)]$. After taking the proper limit,
$E_0=-\lim_{\beta \to +\infty}{{\beta}^{-1}} \log Z_0(\beta)$, one finds,
\begin{equation}
E_0= -{1\over {2\pi}}\int_0^{+\infty}
\log \left(\omega^2\hat f(\omega)\right)\,d\omega.
\label{e0}
\end{equation}
Note that this integral converges thanks to the relation expressed in
\eqn(\ref{deriv}). To be consistent with the perturbative expansion for $E_1$
to come, we can write $E_0$ up to first order in $\phi$,
\begin{equation}
E_0=\frac{\lambda}{2} -{1\over
{4\pi\lambda}}\int_0^{+\infty}(\omega^2+\lambda^2)
\hat\phi(\omega)\,d\omega+O(\phi^2),
\label{e0p}
\end{equation}
the first term being the previously discussed Markovian result, that is the
ground state energy of an harmonic oscillator of frequency $\lambda$. The
computation of $Z_1$ (or $E_1$) is still a formidable task, as the domain of
integration of the Gaussian integral only involves positive $X(\tau)$, for all
$\tau$. The natural impulse is to write ${\cal S}={\cal S}_{osc.}+\delta{\cal
S}$, where ${\cal S}_{osc.}$ is the harmonic oscillator action associated with
a Markovian  process (\eqn{(\ref{Smarkov})}), and,
\begin{eqnarray}
\delta{\cal S}=&-{{1}\over {2}} \int_0^{\beta}
\int_0^{\beta} X({\tau}_1) h({\tau}_1-{\tau}_2) &X({\tau}_2)\,d\tau_1
d\tau_2,\\
=&-\frac{1}{2\beta}\sum_{n=0}^{+\infty} {\hat h}(\omega_n)|\hat X(\omega_n)|^2&,
\end{eqnarray}
where the Fourier transform of $h$ is given in \eqn{(\ref{h})}. $\delta{\cal
S}$ is linear in $\phi$ and can be considered as a small perturbation to
${\cal S}_{osc.}$. We can now use the standard first order cumulant expansion
of quantum mechanics (or statistical field theory), leading to,
\begin{equation}
E_1=\frac{3\lambda}{2} +\lim_{\beta\to +\infty}\langle \delta{\cal S} 
\rangle_{\rm wall} +O(\phi^2),
\end{equation}
where the average is to be taken using the Boltzmann weight associated with
the harmonic oscillator of frequency $\lambda$, with an infinite wall at the
origin. Let us denote by $|\hat l\rangle$ the eigenstates  of this quantum
system (as opposed to $|l\rangle$, the eigenstates of the unconstrained
oscillator), associated with the eigenenergies $\varepsilon_l=((2l+1)+1/2)
\lambda=(2l+3/2)\lambda$ ($l\geq 0$). One can then write,
\begin{equation}
\langle\hat 0 |\hat X(-\omega_n) \hat X(\omega_n)|\hat 0\rangle=
\int_0^{+\infty} 2\cos \omega_n \tau  \sum_{l=0}^{+\infty} |\langle
\hat 0 |X|\hat l \rangle |^2 {\rm e}^{-(\varepsilon_l-\varepsilon_0)\tau}
\,d\tau. 
\label{unit}
\end{equation}
$|\langle \hat 0 |X|\hat l \rangle |^2$ can be computed for the harmonic
oscillator with a wall, using the fact that $\langle x|\hat l \rangle =\sqrt{2}
\langle x| 2l+1 \rangle$, for $x\geq 0$, and exploiting standard properties of
Hermite polynomials. The complete calculation is performed
in appendix A and B. The final result reads,
\begin{equation}
\langle\hat 0 |\hat X(-\omega_n) \hat X(\omega_n)|\hat 0\rangle=
\frac{8}{\lambda^2} \delta \left(\frac{\omega_n}{\lambda} \right)+
\sum_{j=1}^{+\infty} \frac{4jc_j}{4j^2 \lambda^2 + \omega_n^2},
\label{ff2}
\end{equation} 
the Dirac peak coming from the $l=0$ term. The coefficients $c_j$ involved
in this relation read,
\begin{equation}
c_j = \frac{4}{\pi 2^{2j}(2j+1)!} \left(\frac{(2j)!}{j!(2j-1)} \right)^2. 
\end{equation}
Finally, the sum over $n$ in $\delta{\cal S}$ becomes an integral in the
$\beta\to +\infty$ limit, leading to the final expression for $\theta=E_1-E_0$:
\begin{equation}
\theta=\lambda-\frac{1}{2\pi}
\int_0^{+\infty}\hat V(\omega)\hat\phi(\omega)\,d\omega
+O(\phi^2).
\label{f1}
\end{equation}
The kernel $\hat V$ is defined by,
\begin{equation}
\hat V(\omega)=\frac{(\omega^2+\lambda^2)^2}{2\lambda}
\left[\frac{8}{\lambda}\delta\left(\omega\right)
+\sum_{j=1}^{+\infty}\frac{4jc_j}{\omega^2+4j^2 \lambda^2}
-\frac{1}{\omega^2+\lambda^2}\right].
\label{f2}
\end{equation}
As noticed by Oerding {\it et al.}, this cumbersome expression
in terms of Fourier transforms has a remarkably compact form when expressed
in the inverse Fourier space. Indeed, the function between brackets is just
the Fourier transform of,
\begin{equation}
U(\tau)=\frac{1}{\lambda}
\sum_{j=0}^{+\infty}c_j\exp(-2j\lambda|\tau|)-\frac{1}{2\lambda}
\exp(-\lambda|\tau|),
\label{U}
\end{equation}
with $c_0=\frac{4}{\pi}$, so that $\hat V$ is the Fourier transform of
$\frac{1}{2\lambda}(-\partial^2_\tau+\lambda^2)^2U(\tau)$. This allows us to
recast the preceding result into the form,
\begin{equation}
\theta=\lambda-\frac{1}{2\lambda}
\int_0^{+\infty}\phi(\tau)(-\partial^2_\tau+\lambda^2)^2U(\tau)\,d\tau
+O(\phi^2).
\end{equation}
A simple manipulation on the $c_j$'s (see appendix A) allows to resum exactly
the series $(-\partial^2_\tau+\lambda^2)^2U(\tau)$, finally leading to,
\begin{equation}
\theta={ \lambda}\left[1-\frac {2\lambda}{\pi}\int_0^{+\infty}
\phi(\tau)[1-\exp(-2  \lambda\tau)]^{-3/2}\,d\tau\right]+O(\phi^2).
\label{time}
\end{equation} 

We can generalize this expression when the constraint on $X(\tau)$ is
$X(\tau)\geq X_0$, instead of $X(\tau)\geq 0$ \cite{dornic,redner}. Indeed,
for the Brownian walker ($f(\tau)=\exp(-|\tau|/2)$, such that $\langle
X^2\rangle=f(0)=1$), it is known (see section {\bf VII}) that $\theta$
satisfies ${\cal D}_{2\theta}(X_0)=0$ \cite{redner}, where ${\cal
D}_{2\theta}$ is a parabolic cylinder function. We can expand this expression
for small $X_0$, leading to $\theta_{\rm Brownian}
=1/2+X_0/\sqrt{2\pi}+O(X_0)^2$. If we perturb around a general Markovian
process ($f(\tau)=\frac{1}{2\lambda}\exp(-\lambda|\tau|)$), we then get
another perturbative contribution for the exponent $\theta$ (valid in the
limit of small $X_0$), which should be added to the result of
\eqn{(\ref{time})}:
\begin{equation}
\delta\theta(X_0)=2\lambda\times \frac{X_0}{\sqrt{2\pi\langle X^2\rangle}}
+O(X_0)^2
=\lambda^{3/2}\cdot\frac{2X_0}{\sqrt{\pi}}+O(X_0)^2.
\label{x0}
\end{equation}

\section{Resummation: a symmetry argument}

\subsection{Resummation in time}

Consider a Gaussian process of correlator $f$ and persistence exponent $\theta$.
Let us assume that we have been able to resum all terms of the perturbative
expansion which contain only one time integral. Very generally, one can thus
write, 
\begin{equation}
\theta=\int_0^{+\infty} A(f(\tau)/f(0),\tau)\,d\tau.
\label{resum1}
\end{equation}
The variable $f(\tau)/f(0)$ appears due to the fact that $\theta$ should not
depend on the correlator normalization (here $f'(0^\pm)=\mp 1/2$, but
$f(0)=1$ was chosen in \cite{{oerding},{MSBC}}).

If $f(\tau)$ is changed into $f(\alpha\tau)$, it is clear that the persistence
exponent is simply changed into $\alpha\theta$. Using this remark, we get,
\begin{equation}
\alpha\theta=\int_0^{+\infty} A(f(\alpha\tau)/f(0),\tau)\,d\tau,
\end{equation}
which shows after a simple change of variable that, for any process and any
$\alpha$, one must have,
\begin{equation}
 \theta=\int_0^{+\infty} A(f(\tau)/f(0),\tau/\alpha)\,d\tau/\alpha^2.
\label{resum2}
\end{equation}
This strongly suggests that $\theta$ can in fact be written as,
\begin{equation}
 \theta=\int_0^{+\infty} B(f(\tau)/f(0))\,\frac{d\tau}{\tau^2}.
\label{resum3}
\end{equation}
Assuming now that $f(\tau)$ is close to a Markovian process with an associated
small $\phi(\tau)$, one can develop
\eqn{(\ref{resum3})}
leading to,
\begin{equation}
\theta=\int_0^{+\infty} B(\exp(- \lambda\tau))\,\frac{d\tau}{\tau^2}+
\int_0^{+\infty} \phi(\tau)B'(\exp(- \lambda\tau))\,\frac{d\tau}{\tau^2}
+O(\phi^2).
\label{resum4}
\end{equation}
In this perturbative limit, \eqn{(\ref{resum4})} should coincide with 
\eqn{(\ref{time})} leading to,
\begin{equation}
B'(\exp-X)=-\frac 2\pi\frac{X^2}{(1-\exp-2X)^{3/2}},
\label{resum5}
\end{equation}
or, after making the change of variable $u=\exp-X$,
\begin{equation}
B'(u)=-\frac 2\pi\frac{\log(u)^2}{(1-u^2)^{3/2}}
\label{resum6}
\end{equation}
For the integral \eqn{(\ref{resum3})} to converge one should have $B(1)=0$,
finally leading to the final result of \eqn{(\ref{resum3})}, with $B$ given
by:
\begin{equation}
B(u)=\frac 2\pi\int_u^1\frac{\log(v)^2}{(1-v^2)^{3/2}}\,dv.
\label{resum7}
\end{equation}
Note that this expression is not only defined for a nearly Markovian process,
for which $f$ has a cusp at $\tau=0$, but actually converges for any process
for which,
\begin{equation}
f(\tau)/f(0)-1\sim |\tau|^\mu, \quad{\rm when }\quad \tau\to 0,
\end{equation}
for any $\mu>2/3$ (as $B(1-\varepsilon)\sim\varepsilon^{3/2}$).  Smooth
processes (with a continuous velocity) are associated with $\mu=2$, and the
local density of a charge distribution evolving according to the simple
diffusion (or heat) equation corresponds to this case \cite{MSBC}. As we have
argued in detail, $\mu=1$ corresponds to nearly Markovian processes. Finally,
other values of $\mu<2$ correspond to singular walkers for which the fractal
density of the set of $X=0$ crossing times is $1-\mu/2$. Such processes have
been encountered in the study of out of equilibrium atomic surfaces, for which
$X(t)$ is the local height of the substrate \cite{krug}.  

A nice consistency check consists in showing that the first term in
\eqn{(\ref{resum4})} is equal to $\lambda$, that is the Markovian value
for $\theta$. This is simply done by performing an integration by parts using
the explicit expression of $B$, leading to,
\begin{equation}
\theta_{\rm Markov}= \frac{2\lambda}{\pi} \int_0^{+\infty}
\frac{(\lambda\tau)^2{\rm e}^{-\lambda \tau}} {\left(1-{\rm e}^{-2\lambda
\tau}\right)^{3/2}}\cdot\frac{d\tau}{\tau}
=-\frac{2\lambda}{\pi}\int_0^1 \frac{\log u}{\left(1-u^2\right)^{3/2}}\,du
=\frac{2\lambda}{\pi}\int_0^1 \frac{du}{\sqrt{1-u^2}}=\lambda,
\end{equation}
the last integral being obtained through another integration by parts.

The argument presented above was motivated by the following important remark:
for a given correlator $f$, the perturbation $\phi$, or equivalently, the
function $\exp(-\lambda|\tau|)/{2\lambda}$ around which the perturbation
is
started, are actually quite ill-defined. If we knew the complete perturbation
expansion, starting from any value of $\lambda$ we should get the same
result. Note that in standard field theory, one usually perturbs around a
system which is solvable for a certain value (usually 0) of the coupling
constant: there is a unique way of performing the perturbative expansion.
Thus, it is natural to ask whether there is an optimal choice for the starting
value of $\lambda$. A very natural choice is to take for $\lambda$ the value
which cancels the first order perturbative term. In other words, we take the
``best'' starting Markovian correlator such that the first order contribution
vanishes. This gives another non-perturbative expression for $\theta$ (that we
may call ``variational'' or self-consistent perturbative), which must satisfy,
\begin{equation} 
\int_0^{+\infty}
\frac{f(\tau)/f(0)-\exp(-\theta\tau)}{[1-\exp(-2\theta\tau)]^{3/2}}\,d\tau
=0.
\label{np}
\end{equation}
This equation always has a solution, as the expression in \eqn{(\ref{np})} is
clearly positive for $\theta\to+\infty$, and goes to $-\infty$, when $\theta\to
0$. The expression  in \eqn{(\ref{np})} is defined for any $\mu>1/2$, in fact
a larger domain than the fully resummed formula of \eqn{(\ref{resum7})}.

Note that we can write the resummed expression in a similar form,
\begin{equation} 
\int_0^{+\infty}[B(f(\tau)/f(0))-B(\exp(-\theta\tau)]\,d\tau=0,
\end{equation}
which after integration by parts, takes the form,
\begin{equation} 
\int_0^{+\infty}\frac{K[f(\tau)/f(0),\exp(\theta\tau)]}
{\left(1-f^2(\tau)/f^2(0)\right)^{3/2}}\,d\tau=0,
\end{equation}
where the precise form of the known kernel $K$ is of no real interest. This
last remark allows us to make a link with the IIA result. Within this scheme,
based on the approximation that the intervals between the zeros of the process
are independent, it can be shown for smooth processes ($\mu=2$) that $\theta$
must satisfy \cite{MSBC} (with the normalization $f(0)=1$),
\begin{equation} 
1-\frac{\pi\theta}{2\sqrt{-f^{\prime\prime}(0)}}\cdot
\left[1+\frac{2\theta}{\pi}
\int_0^{+\infty}\exp(\theta\tau)\sin^{-1}(f(\tau))\,d\tau\right]=0.
\label{IIA1}
\end{equation}
If one integrates by parts this expression twice, it takes the following form:
\begin{equation} 
\int_0^{+\infty}\exp(\theta\tau)\frac{\left[f^{\prime\prime}(1-f^2)+
ff^{\prime 2} \right](\tau)}
{\left(1-f^2(\tau)\right)^{3/2}}\,d\tau=\sqrt{-2f^{\prime\prime}(0)}.
\label{IIA2}
\end{equation}
This expression now looks of the same type as the ones found within the
perturbative approach. However, its domain of definition remains strictly
$\mu=2$.

Finally, let us mention that when the constraint  on $X(\tau)$ is $X(\tau)\geq
X_0$ \cite{dornic,redner} instead of $X(\tau)\geq 0$, the following
perturbative correction should be added to the preceding expressions for
$\theta$ (see \eqn{(\ref{x0})}):
\begin{equation}
\delta\theta(X_0)=\theta(X_0=0)\cdot \frac{2X_0}{\sqrt{2\pi f(0)}}
+O(X_0)^2.
\label{x0resum}
\end{equation}

\subsection{Resummation in frequency space}

The same argument as above can be applied to the expression of $\theta$ in
frequency space. This time, this will allow us to resum all terms in the
perturbation theory involving only one frequency integration. Again, we assume
that  $\theta$ can be written,
\begin{equation}
\theta=\int_0^{+\infty} C(\hat f(\omega),\omega)\,d\omega.
\label{resumf1}
\end{equation}
We still assume that $f$ has a finite derivative in $0^+$, keeping the
normalization $2|f'(0^+)|=1$.

If $f(\tau)$ is changed into $f(\alpha\tau)/\alpha$ (to preserve the
normalization), $\hat f(\omega)$ is changed into $\hat
f(\omega/\alpha)/\alpha^2$, and
it is again clear that the persistence exponent is simply changed into
$\alpha\theta$. Using this remark, we get:
\begin{equation}
\alpha\theta=\int_0^{+\infty} C\left(
\alpha^{-2} \hat f(\omega/\alpha),\omega\right)\,d\omega,
\end{equation}
which shows after a simple change of variable that, for any process and any
$\alpha$, one must have,
\begin{equation}
 \theta=\int_0^{+\infty} C\left(
\alpha^{-2} \hat f(\omega),\alpha\omega\right)\,d\omega.
\label{resumf2}
\end{equation}
This again strongly suggests that $\theta$ can in fact be written as,
\begin{equation}
 \theta=\int_0^{+\infty} D(\omega^{2}\hat f(\omega))\,d\omega.
\label{resumf3}
\end{equation}
Note that this property is shared by the exact and general expression of
$E_0$, given in \eqn(\ref{e0}). 

Now, one can consider a nearly Markovian process, for which the correlator
satisfies \eqn(\ref{h}). One can develop \eqn(\ref{resumf3}) up to first order
in $\hat\phi$ and identify the result to the perturbation result of
\eqn(\ref{f1}-\ref{f2}). The calculation is elementary and leads to,
\begin{eqnarray}
\theta=\frac 4\pi {\hat f^{-1/2}(0)}+{1\over {2\pi}}
\int_0^{+\infty}\hat W(\omega^2\hat f(\omega))\,d\omega,\\
\hat W(x)=\sum_{n=1}^{+\infty}\frac{c_n}{n}\log\left(1+4n^2
\left(x^{-1}-1\right)\right)+\log(x).
\label{eight}
\end{eqnarray}
The first term arises from the $\delta$ term in the kernel $\hat V$, and can
also be written as $\frac 8\pi\int_0^{+\infty}\delta\left(\omega|\hat
f(\omega)|^{1/2}\right)\,d\omega$. Again, it is easy using relations given in
appendix A to check that the Markovian value $\theta=\lambda$ is recovered for
the Markovian correlator $\hat f(\omega)=(\omega^2+\lambda^2)^{-1}$. Note
finally that this procedure permits the recovery of the exact expression of
$E_0$, which is produced by the last $\log(x)$ term in the kernel $\hat W$.

\section{Perturbation around a non-Gaussian Markovian process}

When writing ${\cal S}={\cal S}_{osc.}+\delta{\cal S}$, we  deliberately
chose to perturb around a Gaussian Markovian walker, or around the quantum
action of an harmonic oscillator in terms of the pseudo-action ${\cal S}$.
This is quite arbitrary, and in principle the action of any (preferably
solvable) quantum system would have worked. The stochastic process associated
to such an action (each trajectory $\{X(\tau)\}$ being weighted by $\exp-{\cal
S}[\{X(\tau)\}]$) is Markovian but, in general, non-Gaussian, as the only
Gaussian quantum action is that of an harmonic oscillator.

So in this section, we consider a stationary stochastic process $X(\tau)$ of
any kind, associated with the weight or  pseudo-action ${\cal S}$, and a
quantum mechanical system for which the action is ${\cal S}_{\rm Q}$ (``Q''
for ``quantum''). This quantum system could be an harmonic oscillator, a
particle in a square box, and more widely, any system preferably solvable for
the actual perturbative calculation to be tractable.

Then, setting ${\cal S}={\cal S}_{\rm Q}+ \delta{\cal S}$, and reproducing
exactly the calculations of the beginning of section {\bf IV}, we end up with
the following perturbative expression for $\theta$:
\begin{equation}
\theta=E_1^{\rm Q}-E_0^{\rm Q} +\lim_{\beta\to +\infty}[\langle \delta{\cal S} 
\rangle_{1}-\langle \delta{\cal S}\rangle_0] +O(\delta{\cal S}^2),
\label{gen}
\end{equation}
where $E_0^{\rm Q}$ (respectively $E_1^{\rm Q}$) is the ground state energy of
the unconstrained (respectively with an infinite wall at $X=0$) quantum
system.  $\langle \quad \rangle_{0}$ and $\langle \quad\rangle_1$ denote
quantum averages performed using the Hamiltonian of the quantum system,
respectively without and with the infinite barrier at the origin.

We have already implicitly used \eqn{(\ref{gen})} in section {\bf IV}, where
${\cal S}_Q$ was chosen to be the Gaussian quantum action of an harmonic
oscillator. Let us now illustrate \eqn(\ref{gen}) by taking a non-Gaussian
system as the starting quantum system. The simplest possible example is that
of particle in a box, with $X$ restrained to the interval $[-b,b]$. We now use
this simple non-Gaussian system to compute approximately the value of $\theta$
for a Gaussian process associated with the Gaussian weight ${\cal S}$ defined
in \eqn{(\ref{weight})}.

Let us call $|l\rangle$ ($l \geq 0$), the eigenstates of a quantum particle in
the box $[-b,b]$, associated with the eigenenergies
$\varepsilon_l=\frac{1}{2}k_0^2 (l+1)^2$, with $k_0=\frac{\pi}{2b}$. The
eigenstates of the constrained system (a particle in the box $[0,b]$) are the
$|\hat l\rangle=\sqrt{2}|2l+1\rangle$ ($l \geq 0$). To evaluate $\langle
\delta{\cal S}  \rangle_{1}-\langle \delta{\cal S}\rangle_0$, one essentially
needs to compute $\langle 0|X(-\omega_n)X(\omega_n)|0\rangle $ and $\langle
\hat 0|X(-\omega_n)X(\omega_n)|\hat 0\rangle $. This is a straightforward task
using identities similar to \eqn{(\ref{unit})}, where the scalar products,
$\langle 0|X|l\rangle$ and $\langle \hat 0|X|\hat l\rangle$ are even easier to
compute for a particle in a box (see appendix B). Introducing,
\begin{equation}
\hat k(\omega_n)=\langle 0|X(-\omega_n)X(\omega_n)|0\rangle =
\frac{256}{\pi^2} \sum_{j=0}^{+\infty}\frac{a_j b_j}{k_0^4 b_j^2 +
\omega_n^2},
\label{k}
\end{equation}
with,
\begin{equation}
a_j = \frac{2(j+1)^2}{(2j+1)^4(2j+3)^4}, \quad {\rm and}\quad b_j =
\frac{1}{2}(2j+1)(2j+3), 
\end{equation}
we get the following expressions for $E_1$ and $E_0$ ($\theta=E_1-E_0$):
\begin{equation}
E_0 = \frac{k_0^2}{2} + \frac{1}{2\pi} \int_0^{+\infty}  \left(
\frac{\hat k(\omega)}{\hat f(\omega)}  - 1 \right)\,d\omega,
\label{box1}
\end{equation}
and,
\begin{eqnarray} E_1 & = & 2 k_0^2 + \frac{\pi^2}{32} k_0^2 \left(
\frac{1}{k_0^4\hat f(0)} -\frac{12}{15-\pi^2} \right) \\ 
& + & \frac{1}{32\pi} 
\int_0^{+\infty}   \left( \frac{1}{\hat f(\omega)} -
\frac{1}{\hat k(\omega)} \right) \hat k(\omega/4) \, d\omega.
\label{box2}
\end{eqnarray} 

For a given correlator $\hat f(\omega)$, it is not clear what  the ``best''
starting value for $k_0$ (or for the box size $b$) is. Let us propose two
natural choices. We can take $k_0$ such that the first order perturbation
vanishes, which leads to $\theta=3k^2_0/2$. An alternative choice is to take
$k_0$ such that $E_1$ is minimum, as it can be shown that $E_1(k_0)$ has
always such a minimum for a finite $k_0$. In fact, the variational inequality
$E_1\leq 2k_0^2 +\lim_{\beta\to +\infty}\langle \delta{\cal S}  \rangle_{1}$
is exact for any $k_0$, which intuitively validates this choice if $k_0$.

\section{Generalized persistence}

So far, we have essentially considered the probability that the signal
$X(\tau)$ has never changed sign. In fact, it seems natural to study the more
general probability that the signal has always remained above a certain level
$X_0$. When $X_0\neq 0$, this defines the $X_0$-level persistence. This
generalized persistence has already been introduced for the simplest Markovian
Gaussian walker \cite{redner}, and spin systems \cite{dornic}.
Moreover, at least in the framework of Mazenko approximation \cite{M1}, there
is a connection between the persistence of the $q$-Potts model
\cite{{BD1},{D1},{BD2}} (the probability that a given site always remains in a
given phase) and the $X_0=F(q)$-level persistence of a certain Gaussian
variable \cite{SM}. 

Let us take the example of the Gaussian Markovian walker, associated, within
our formalism, with the action of an harmonic oscillator. $E_1(X_0)$ is now the
ground state energy of an harmonic oscillator with an infinite barrier at
$X_0$, for which the eigenstates can be expressed in terms of a parabolic
cylinder function \cite{redner} (generalization to a continuous index of
Hermite polynomials). $E_1$ is then implicitly defined by imposing that the
ground state eigenfunction has a unique node at $X=X_0$. If we come back to
real time $t=\exp\tau$ (see section {\bf III}), $X_0$-level persistence for
the Langevin walker satisfying $\frac{dx}{dt}=\eta(t)$ is defined as the
probability that $x(t)$ always remained greater than $X_0\sqrt{\langle
x^2(t)\rangle }=X_0\sqrt{t}$. This decays as a power-law of time with exponent
$\theta(X_0)=E_1(X_0)-E_0$.

If we were to compute the $X_0$-level persistence exponent $\theta(X_0)$ for a
Gaussian process using the perturbation theory formalism, we would have to
evaluate scalar products like $\langle \hat 0|X|\hat l\rangle$, where $|\hat
l\rangle$ are the eigenstates of the harmonic oscillator with an infinite
barrier at $X=X_0$. Unfortunately, this seems to be an analytically untractable 
problem. However, using the formalism of the preceding section, we only have
to evaluate brackets involving eigenstates of a particle in the box $[X_0,b]$,
which are explicitly known. The calculation is straightforward (see appendix
B) and leads to,
\begin{eqnarray} E_1(X_0) & = & \frac{2 k_0^2}{(1-\eta)^2} + 
\frac{\pi^2}{32}k_0^2
(1+\eta)^2\left(
\frac{1}{k_0^4\hat f(0)} -\frac{12}{15-\pi^2} \right) \\ 
& + & \frac{k_0^2(1-\eta)^4}{32\pi} 
\int_0^{+\infty}   \left( \frac{1}{\hat f(\omega)} -
\frac{1}{\hat k(\omega)} \right) \hat k(\omega(1-\eta)^2/4) \, d\omega,
\label{box3}
\end{eqnarray} 
where $\eta=X_0/b$, and $\hat k$ has been defined in \eqn{(\ref{k})}. As a
check, we can see that for $\eta = 0$ ($i.e.$ the wall is at the origin) we
recover the result of \eqn{(\ref{box2})}, and for $\eta = -1$ ($i.e.$ the
wall is at  $X_0 =  -b $, which corresponds to no effective constraint), we
recover the expression for $E_0$ of \eqn{(\ref{box1})}.

Again, for a given process $X$ and a given level $X_0$, $k_0$ can be fixed by
imposing that the first order perturbation term in $\theta$ vanishes, or by
taking the value of $k_0$ for which $E_1(X_0,k_0)$ is minimum.

\section{Numerical simulations}

We now illustrate the various analytical results obtained in the
preceding sections by means of numerical simulations.

\subsection{Nearly Markovian processes}

As already mentioned in the introduction, a local Ising spin evolving after a
quench at $T=0$, from the high temperature disordered phase, essentially
behaves as the sign of a Gaussian variable. Mazenko approximation \cite{M1}
then permits the calculation of the two-time correlator of this Gaussian
process. It happens that, in one dimension, this approximation recovers the
exact expression of $\langle S(t)S(t')\rangle$ \cite{AB}, leading to the
following form of the correlator $f$ when expressed in the fictitious time
$\tau=\log(t)$:
\begin{equation}
f(\tau)=\sqrt{\frac{2}{1+\exp(\tau)}}.
\label{ising}
\end{equation} The exact value of $\theta$ in $d=1$ is $\theta=3/8=0.375$
\cite{AB1}. The ``variational'' and resummed perturbative expression of
\eqn{(\ref{np})} and \eqn{(\ref{resum3}-\ref{resum7})} respectively lead to
$\theta_{\rm var}=.3595\ldots$ and $\theta_{\rm pert}=.3677\ldots$. The
process associated with the correlator given by \eqn{(\ref{ising})} has been
actually simulated using the Fourier space form of \eqn{(\ref{motion})}. We
have obtained $\theta=0.355\pm 0.005$, in extremely good agreement with the
theory. The small discrepancy with the exact result 0.375 for the Ising model
is attributed to the fact that the actual process such that $S(t)={\rm
sign}(X(t))$ is not strictly Gaussian. However it seems that this
non-Gaussian effect is rather small.

We have also tested our theoretical expressions using a correlator introduced
in \cite{pert}:
\begin{equation}
f(\tau)=\frac{2}{5}\exp(-\tau)+\frac{3}{5}\exp(-2\tau).
\label{march}
\end{equation}
We found $\theta_{\rm var}=1.4855\ldots$ and $\theta_{\rm pert}=1.4802\ldots$,
again in good agreement with the numerical result $\theta=1.481\pm 0.005$.

\subsection{Other singular processes}

Interesting examples of singular correlators with $\mu<2$ and $\mu\neq 1$ (see
the definition of $\mu$ in section {\bf V}) have been introduced in the
framework of dynamical surfaces described by the following time-dependent
equation \cite{krug}:
\begin{equation}
\label{Langevin}
\frac{\partial h}{\partial t} = - (- \nabla^2)^{z/2} h + \eta,
\end{equation}
where $h$ is the local height of the fluctuating interface, and $\eta$ is
Gaussian white noise. The equation being linear, $h(x,t)$ is a Gaussian
variable, for which we take the initial condition $h(x,0) = 0$.
To define the first passage problems of interest, consider
the quantity,
\begin{equation}
\label{P(t_0,t)}
P(t_0,t) = 
{\rm Prob}[h(x,s) \neq h(x,t_0) \;\; \forall s : t_0 < s < t_0 + t],
\end{equation}
and define $\theta_0$ and $\theta_s$ as,
\begin{eqnarray}
p_0(t) &\equiv P(0,t) \sim t^{-\theta_0}, \;\;\; t \to +\infty,&\\
p_s(t) &\equiv \lim_{t_0 \to +\infty}  P(t_0,t) \sim t^{-\theta_S}, 
\;\;&\; t \to +\infty.
\end{eqnarray}
$p_0$ measures the first passage exponent of the growing interface, whereas
$p_s$ contains the relevant information, when the interface has entered the
steady state ($t_0\to+\infty$).

The correlators associated with these two persistence problems are respectively
(when expressed as functions of the fictitious time $\tau$)\cite{krug}:
\begin{eqnarray}
f_0(T) &= \cosh(\tau/2)^{\mu} - \vert \sinh(\tau/2) \vert^{\mu}&\\
f_s(T) &= \cosh(\mu \tau/2) - 
\frac{1}{2}\vert 2 \sinh(\tau/&2) \vert^{\mu},
\end{eqnarray}
and both satisfy $1-f_{0,s}(\tau)\sim\tau^\mu$, for small $\tau$, with
$\mu=1-d/z$ ($\mu=1-(d+2)/z$ for a volume conserving noise).
We now simply treat $\mu$ as a free parameter. Using a connection to the
fractional Brownian walker, it has been conjectured that $\theta_s=1-\mu/2$
\cite{krug}, which has been confirmed by numerical simulations.

Let us take two typical values for $\mu$. For $\mu=3/2$, we find 
$\theta_{\rm 0,var}=0.2088\ldots$ and $\theta_{\rm 0,pert}=0.2146\ldots$,
which compare well to the numerical value $\theta_0=0.201\pm 0.005$. For the
case of the steady interface, the conjectured persistence exponent is
$\theta_s=1/4$, in good agreement with the simulations ($\theta_s=0.247\pm
0.005$). Variational and perturbative methods are reasonably accurate, giving
$\theta_{\rm s,var}=0.2583\ldots$ and $\theta_{\rm s,pert}=0.2644\ldots$. Note
that the first order perturbation  expression of \eqn{(\ref{time})} reproduces
exactly the conjectured value for $\theta_s$.

We have also tested the case $\mu=3/4$, which is getting dangerously close to
the limit of the validity domain of our variational ($\mu_{\rm var}=1/2$) and
perturbative ($\mu_{\rm pert}=2/3$) expressions. The numerical value of
$\theta_0$ is $\theta_0=0.85\pm 0.01$, for which the variational approach
gives $\theta_{\rm 0,var}=0.8852\ldots$. Not surprizingly, the resummed
perturbation leads to a bad result ($\theta_{\rm 0,pert}\approx 1.1$). The
conjectured value for $\theta_s$ is $\theta_s=0.625$, while the simulation of
the process leads to $\theta_s=0.625\pm 0.005$, and that of the associated
discrete solid-on-solid model leads to $\theta_s=0.635\pm 0.005$ (see
\cite{krug} for details). We find a qualitatively correct value of
$\theta_{\rm s,var}=0.6662\ldots$, but the resummed perturbation fails again
($\theta_{\rm s,pert}\approx 0.84$).

\subsection{Smooth processes}

For singular processes ($\mu<2$), it was not possible to compare our
variational and perturbative results to the IIA expressions of 
\eqn{(\ref{IIA1}-\ref{IIA2})}, which are only defined for smooth processes.

One of the most spectacular example of smooth Gaussian processes has been
given in \cite{MSBC}: consider an initially random spatial distribution of
charges of zero average, $\rho({\bf x},t=0)$. It then evolves according to the
simple diffusion equation,
\begin{equation}
\frac{\partial\rho}{\partial t}({\bf x},t)=\nabla^2\rho({\bf x},t).
\end{equation}
The persistence is defined as the probability that the local charge at a
given ${\bf x}$ never changes sign. It decays as a power-law, defining
$\theta_d$, the dimension-dependent persistence exponent.

The IIA \cite{MSBC} (as well as the specific method of \cite{newman}) is in
amazing agreement with numerical simulations. For instance, in $d=1$,
$\theta_{\rm IIA}=0.1203\ldots$, to be compared to the numerical value
$\theta=0.1207\pm 0.0005$. The agreement seems to be of the same order in any
dimension. Smooth processes are in principle beyond the range of application
of perturbative methods. Still, the variational approach remains
qualitatively correct, leading to $\theta_{\rm var}=0.1428\ldots$ for the
one-dimensional diffusion equation, whereas the resummed perturbation theory is
again quite bad ($\theta_{\rm pert}=0.1612\ldots$).

Another example of smooth process is the Gaussian walker satisfying
$\frac{d^nX}{dt^n}=\eta(t)$, for $n\geq 2$ ($n=1$ being the Markovian Brownian
walker which is singular). The case $n=2$ corresponds to a particle submitted
to a random force, for which the persistence exponent is known to be
$\theta=1/4$ \cite{theta2}. The two-time correlator when expressed in the
fictitious time reads,
\begin{equation}
f(\tau)=\frac{3}{2}\exp(-|\tau|/2)-\frac{1}{2}\exp(-3|\tau|/2).
\end{equation}
The IIA leads to $\theta_{\rm IIA}=0.2647\ldots$, whereas $\theta_{\rm
var}=0.2857\ldots$, and $\theta_{\rm pert}=0.3198\ldots$.

\subsection{Perturbation around the action of a particle in a box}

Let us briefly give a few applications of our expressions of
\eqn{(\ref{box1},\ref{box2},\ref{box3})}.

As a simple test, they have been applied to the case of the Markovian walker
with $\lambda=\theta=1/2$, and $X_0=0$. The ``variational'' approach, which
consists in taking the size of the box (or $k_0$) such that the first order
perturbation vanishes leads to $2\theta_{\rm var}=1.0074\ldots$. For the
correlator given by \eqn{(\ref{march})}, we found $\theta_{\rm
var}=1.4323\ldots$, in fair agreement with simulations and perturbative
approaches around a Markovian process. 

Finally, we have tested \eqn{(\ref{box3})} in the case of the Brownian walker
(for which $\lambda=1/2$), for $X_0 \neq 0$. In this case, it
is known that $\theta$ satisfies ${\cal D}_{2\theta}(X_0)=0$
\cite{redner}, where ${\cal D}_{2\theta}$ is a parabolic cylinder function.
For $X_0=1/3$, (comparable to $\langle X^2\rangle=f(0)=1$, for $\lambda=1/2$),
we found $\theta_{\rm var}=0.7032\ldots$, to be compared to the exact value
$\theta=0.6440\ldots$. Note that the perturbative expression of
\eqn{(\ref{x0resum})} leads to $\theta=0.6330\ldots$.

\section{Conclusion} 

In this paper, we have stressed the importance of studying  persistence for
Gaussian stationary processes, as the calculation of $\theta$ for many
physical systems can be often mapped on the persistence problem for this kind
of process. We have then extended the perturbative approach around a Markovian
process, introduced in \cite{pert}. We have obtained  resummed perturbative
expressions (\eqn{(\ref{resum4},\ref{resum7},\ref{eight})}) and a new
self-consistent perturbative (or variational) expression for the persistence
exponent (\eqn{(\ref{np})}). It seems that this variational result is more
effective in reproducing numerical results, sometimes with impressive
accuracy. We have also shown that all these expressions take a similar form as
the alternative result of the IIA, which only applies to smooth processes. We
have also given perturbative expressions for the $X_0$-level persistence
exponent (\eqn{(\ref{x0},\ref{x0resum})}). Finally, we have shown that this
type of perturbative approach is even more general, as the starting process
around which we decide to perturb can be any Markovian process associated with
a (preferably solvable) quantum problem. We have illustrated this point by
explicitly deriving a variational expression for the $X_0$-level persistence
exponent, when the starting quantum system is chosen to be a particle in a
bounded box (\eqn{(\ref{box1},\ref{box2},\ref{box3})}).

Finally, we conclude by pointing out that our perturbative and variational
techniques have been useful in a wide variety of problems. This includes the
calculation of the survival probability of a mobile particle in a fluctuating
field \cite{MC}, and the calculation of global persistence exponent in
critical spin systems (to compute the order $\varepsilon^2=(4-d)^2$
perturbative correction \cite{oerding}), and for directed percolation
\cite{OF}. 

\acknowledgments
We are grateful to A.J. Bray, S.J. Cornell and J. Krug for numerous
discussions since we all started working on persistence. J. Basson and D. Dean
are warmly thanked for their wise comments on the original manuscript.

\newpage
\vskip 0.9cm
{\centerline {\bf APPENDIX A}}
\vskip 0.6cm

We want to compute $c_j = |\langle \hat 0|x|\hat j\rangle|^2$, where 
$|\hat j\rangle$ is the $j$th eigenstate of the harmonic oscillator with an
infinite barrier at the origin.
\begin{equation}
\langle \hat j| x \rangle = \sqrt{2}\langle 2j+1| x \rangle = 
\frac{\sqrt{2}}{\sqrt{2^{2j+1}(2j+1)! \sqrt{\pi}}}
H_{2j+1}(x) {\rm e}^{-\frac{x^2}{2}}.
\end{equation}
The extra factor $\sqrt{2}$ is due to the fact that $\langle \hat j| x
\rangle$ is only defined on the interval $[0;+\infty]$, but should still be
normalized. One then finds, 
\begin{equation}
c_j = \frac{4}{\pi 2^{2j} (2j+1)!} I_j^2,
\end{equation} 
where $I_j = \int_0^{+\infty}x^2 {\rm e}^{-x^2} H_{2j+1}(x)\,dx $ can be readily
calculated, using the properties $H_{n+1}(x) = 2xH_n(x) - 2nH_{n-1}(x)$,
$H'_n(x) = 2nH_{n-1}(x)$ and $\int_0^{+\infty} {\rm e}^{-x^2} H_n(x)\,dx =
H_{n-1}(0)$. This yields, 
\begin{equation}
I_j = H_{2j}(0) + 4jH_{2j-2}(0) = (-1)^{j+1} \frac{(2j)!}{j!(2j-1)},
\end{equation}
which finally gives the result of the main text.

The $c_j$'s satisfy the recursion relation,
\begin{equation}
\frac{c_{j+1}}{c_j} = \frac{(2j-1)^2}{(2j+3)(2j+2)},
\label{cj}
\end{equation}
which shows that the generating function $f(x) = \sum_{j=0}^{+\infty} c_j x^j$
satisfies the  hypergeometric differential equation $x(1-x)f'' + \frac{3}{2} f'
- \frac{1}{4}f = 0$. The (unique) solution with $f(0) =c_0= \frac{4}{\pi}$ is
given by 
\begin{equation}
f(x) = \frac{4}{\pi} F\left(-\frac{1}{2},-\frac{1}{2},\frac{3}{2},x\right), 
\end{equation}
which yields the identities $\sum_{j=0}^{+\infty} c_j = f(1) = \frac{3}{2}$ 
and $\sum_{j=0}^{+\infty} jc_j = f'(1) = \frac{1}{4}$. 

Now defining $U(\tau)$ as in \eqn{(\ref{U})} we get,
\begin{equation}
\frac{1}{2\lambda}
(-\partial^2_\tau+\lambda^2)^2U(\tau)=
\frac{\lambda^2}{2}\sum_{j=0}^{+\infty}C_j\exp(-2j\lambda|\tau|)=
\frac{\lambda^2}{2}S[\exp(-2\lambda|\tau|)],
\label{dU}
\end{equation}
where $C_j=(4j^2-1)^2c_j$ satisfies (using \eqn{(\ref{cj})}),
\begin{equation}
\frac{C_{j+1}}{C_j} = \frac{j+\frac{3}{2}}{j+1}.
\label{CCj}
\end{equation}
We recognize the recursion relation obeyed by the coefficients of the series
expansion of the function $S(x)=\frac{4}{\pi}(1-x)^{-3/2}$, which leads to the
final result of \eqn{(\ref{time})}.

\newpage
{\centerline {\bf APPENDIX B}}
\vskip 0.6cm

In this appendix, we write the general equation for $E_0$ and $E_1$, when the
perturbation theory is applied to ${\cal S}={\cal S}_{\rm Q}+\delta{\cal S}$.
${\cal S}_{\rm Q}$ is assumed to be the action associated with the  quantum
Hamiltonian $H_0$, with eigenenergies $\varepsilon_l$ and eigenstates
$|l\rangle$. The associated Hamiltonian with an infinite wall at the origin is
$H_1$, with eigenenergies $\hat \varepsilon_l$ and eigenstates $|\hat
l\rangle$. When the potential is symmetric with respect to $X=0$, one simply
has  $\langle x|\hat l\rangle=\sqrt{2}\langle x|2l+1\rangle$, and $\hat
\varepsilon_l=\varepsilon_{2l+1}$.

\subsection{Equation for $E_0$}
The lower ``energy'' $E_0$ is a functional of  the inverse correlator $\hat
g(\omega) = 1/\hat f(\omega)$: 
\begin{equation}
E_0 = \varepsilon_0  + \lim_{\beta\to +\infty}
\frac{1}{2\beta} \sum_ {n=0}^{+\infty}(\hat g(\omega) -
\hat g_0(\omega)) \cdot \langle 0|X(-\omega_n)X(\omega_n)|0\rangle, 
\end{equation}
where,
\begin{eqnarray}
\hat k(\omega_n)&=\hat g_0^{-1}(\omega_n) = 
\langle 0|X(-\omega_n)X(\omega_n)|0\rangle,\qquad & 
\label{cod}\\
&= \int_0^{+\infty} 2\cos
\omega_n \tau  \sum_{l=0}^{+\infty} |\langle 0 |x|l \rangle |^2 
{\rm e}^{-(\varepsilon_l -{\varepsilon_0 })\tau}& {\rm d} \tau  =
\sum_{l=1}^{+\infty}
\frac{2d_l  m_l^2}{d_l^2 + \omega_n^2},  
\end{eqnarray}
with $d_l = \varepsilon_l  - \varepsilon_0 $, $m_l = |\langle 0 | x|l
\rangle|$ ($m_0=0$ due to the symmetry of the potential). Note that $\hat k$
is nothing more than the Fourier transform of the two-time correlation function
of the position of the considered quantum particle.

Then, transforming the sum over Matsubara frequencies into an
integral, one obtains:
\begin{equation}
E_0 = \varepsilon_0  + \frac{1}{2\pi} \int_0^{+\infty}  \left(
\frac{\hat k(x)}{\hat f(x)}- 1
\right)\, dx. 
\end{equation}

Due to the sum rule $\sum_{l=1}^{+\infty} 2 d_l m_l^2 = 1$, valid for any
Hamiltonian,  the integrand tends to 0 as $x \to +\infty$. 

In the text, we have considered two examples for the starting quantum
correlator $\hat k$: 
\begin{itemize}
\item Harmonic oscillator of frequency $\lambda$:  \\ $\varepsilon_l  =
\lambda(l+\frac{1}{2})$, and $d_l = \lambda l$;  $m_l = (2\lambda)^{-1/2}
\delta_{l,1}$;  $\hat k(\omega) = (\omega^2 + \lambda^2)^{-1}$, which directly
leads to \eqn{(\ref{e0p})}, obtained in section {\bf IV} by expanding the
exact result of \eqn{(\ref{e0})}.

\item Particle in a box of width $2b =\pi/k_0$: \\ $\varepsilon_l  =
\frac{1}{2}k_0^2(l+1)^2$, and $d_l = \frac{1}{2} k_0^2 l(l+2)$. After an
elementary calculation involving the eigenstates of a particle in a box, we
get, 
\begin{equation}
m_l = (1-(-1)^l)\frac{1}{\pi k_0} \frac{4(l+1)}{l^2(l+2)^2}, \quad \hat
k(\omega)
= \frac{256}{\pi^2} \sum_{j=0}^{+\infty}\frac{a_j b_j}{k_0^4 b_j^2 +
\omega^2}, 
\label{box5}
\end{equation} 
with, 
\begin{equation}
a_j = \frac{2(j+1)^2}{(2j+1)^4(2j+3)^4}, \quad b_j = \frac{1}{2}(2j+1)(2j+3). 
\end{equation} 
We then recover the expression of  \eqn{(\ref{box1})}.

Note that the following sum rules have been used in the main text:
\begin{equation}
\frac{256}{\pi^2}\sum_{j=0}^{+\infty}a_j b_j=1,\quad
\frac{256}{\pi^2}\sum_{j=0}^{+\infty}a_j b_j^2=\frac{1}{2},\quad
\frac{256}{\pi^2}\sum_{j=0}^{+\infty}\frac{a_j}{b_j}=
\frac{5}{4}-\frac{\pi^2}{12}.
\end{equation}
\end{itemize}

\subsection{Equation for $E_1$}

The ``energy'' $E_1$ is also a functional of 
the inverse correlator $\hat g(\omega) = 1/\hat f(\omega)$: 
\begin{equation}
E_1 = \hat\varepsilon_0  + \lim_{\beta\to +\infty} \frac{1}{2\beta}
\sum_{n=0}^{+\infty} (\hat g(\omega_n) - \hat g_0(\omega_n)) \cdot \langle
\hat 0|X(-\omega_n)X(\omega_n)| \hat 0\rangle, 
\end{equation}
where, $\hat k=\hat g_0^{-1}$ has been defined in  \eqn{(\ref{cod})}. Let us
now introduce,
\begin{eqnarray} 
\hat K(\omega_n)&=\langle \hat 0|X(-\omega_n)X(\omega_n)|\hat 0\rangle,& \\
&= \int_0^{+\infty} 2\cos
\omega_n \tau  \sum_{l=0}^{+\infty} |\langle \hat 0 &|x|\hat l \rangle |^2 
{\rm e}^{-(\hat\varepsilon_l -{\hat\varepsilon_0 })\tau} {\rm d} \tau  = 
\sum_{l=0}^{+\infty} \frac{2\hat d_l  \hat m_l^2}{\hat d_l^2 + \omega_n^2}.  
\end{eqnarray}
$\hat K$ is the two-time correlator of the position of the quantum particle in
the presence of the wall. As before, $\hat d_l = \hat\varepsilon_l  -
\hat\varepsilon_0 $,  $\hat m_l = |\langle \hat 0 | x| \hat l\rangle|$.  Note
that, contrary to the calculation for $E_0$, the $l=0$ contribution in the sum
above is non-zero so that, strictly speaking, this term should be written
$2\pi\hat m_0^2\delta(\omega_n)$. This term has been written under this form
in the main text.

Finally, the general expression of $E_1$ reads,
\begin{equation}
E_1 = \hat \varepsilon_0  + \frac{1}{2\pi} \int_0^{+\infty}   \left(
\frac{1}{\hat f(x)} - \frac{1}{\hat k(x)}
\right)\hat K(x)\, dx. 
\end{equation}
We can again make this result more explicit for both considered quantum
systems:
\begin{itemize}
\item Harmonic oscillator of frequency $\lambda$:  \\ $\hat\varepsilon_l  =
\lambda(2l+\frac{3}{2})$,  $\hat d_l = 2\lambda l$, and  $\hat m_l =
\sqrt{c_l}$, which  leads to the expressions of \eqn{(\ref{ff2}-\ref{f2})}.

\item Particle in a box of width $2b =\pi/k_0$:\\
In this case, $\hat K$ is the Fourier transform of the two-time  correlator of
the position of a particle in a box of size $b =\pi/2k_0$. It is then clear
using \eqn{(\ref{box5})} that,
\begin{equation}
\hat K(\omega)=\frac{1}{16}\hat k(\omega/4)+\frac{\pi^3}{8k_0^2}
\delta\left(\omega\right),  
\end{equation}
the extra Dirac peak coming from the fact that the operator $X$ now has a finite
average, as the particle belongs to the interval $[0,b]$. This immediately
leads to the formula of \eqn{(\ref{box2})}, using $\hat
k(0)=\frac{15-\pi^2}{12k_0^4}$.

When the constraint is $X\geq X_0$, with $X_0=\eta b$ (as in section {\bf
VII}), the quantum particle now lives in a box of size $(1-\eta)b$, and the
expression for $\hat K$ is changed accordingly, leading to,
\begin{equation}
\hat K(\omega)=
\frac{(1-\eta)^4}{16}\hat k\left(\omega\frac{(1-\eta)^2}{4}\right)
+\frac{\pi^3}{8k_0^2}(1+\eta)^2
\delta\left(\omega\right).  
\end{equation}
This immediately leads to the result of \eqn{(\ref{box3})}.
\end{itemize}

\end{document}